\documentclass[12pt]{article}\usepackage{amsmath,amsfonts}
\makeatletter \@addtoreset{equation}{section}

\makeatother
\begin{document}
\baselineskip 18pt%

\begin{titlepage}
\vspace*{1mm}%
\hfill%
\vbox{
    \halign{#\hfil        \cr
           IPM/P-2006/053 \cr
           SUT-P-06-11b   \cr 
                     } 
      }  
\vspace*{15mm}%

\centerline{{\Large {\bf  Double-Horizon Limit and Decoupling of the
  }}} \centerline{{\Large {\bf Dynamics at the Horizon }}}

\vspace*{5mm}
\begin{center}
{ H. Arfaei, R. Fareghbal}%

\vspace*{0.8cm}{\it {Institute for Studies in Theoretical Physics and Mathematics (IPM)\\
P.O.Box 19395-5531, Tehran, IRAN}}\\
{\it {And}}\\
{\it {Department of Physics, Sharif University of Technology\\
P.O.Box 11365-9161, Tehran, IRAN}}\\

{E-mails: {\tt arfaei@ipm.ir, fareghbal@theory.ipm.ac.ir}}%
\vspace*{1.5cm}
\end{center}

\begin{center}{\bf Abstract}\end{center}
\begin{quote}

We show that the \emph{ attractor mechanism} for generic black hole
is a consequence of the double-horizon.  Investigation of equations
of motion shows that in the case of the double-horizon black
 holes, the dynamics of the geometry, the scalars and the gauge fields at the
 horizon decouples from the rest  of the space.
 In the general case,  the value of the fields at the horizon satisfies a
 number of differential equations of functions of the $\theta$ coordinate.
 We show this for the case of rotating and non-rotating electrically
 charged black holes in the general two derivative theories of gravity
  and $f(R)$ gravities including the  theories with cosmological constant.

\end{quote}
\end{titlepage}
\section{Introduction }
The attractor mechanism for black holes states that the value of the
scalar fields (moduli fields  ) at the horizon,  in certain class of
black holes (extremal black holes),  is fixed and is independent of
the large distance boundary conditions \cite{Ferrara:1995ih}.

 It was first shown for black holes in  supersymmetric theories but
recent careful analysis  has confirmed it for non-supersymmetric
case \cite{Goldstein:2005hq}-\cite{Kallosh:2006bt}.  This
generalization is  studied by several works from different points of
view. \cite{Tripathy:2005qp}-\cite{Ferrara:2006xx}.

Almost at the same time of this progress, Sen has developed his
entropy function method \cite{Sen:2005wa}-\cite{Sen:2005iz}. In
this method  the $AdS_2\times S^2$ geometry of the near horizon
geometry of the spherically symmetric extremal black holes plays
the central role and  the parameters of the near horizon  are
fixed by extremizing a function called  the \emph{entropy
function}. Charges of the black hole play main role in fixing the
parameters of the near horizon geometry and including  the values
of the moduli. Therefore all the relevant  information about
horizon is fixed independent of the asymptotic behavior. Following
this progress, a number of articles  have appeared in studying
various aspects of entropy function method
\cite{Alishahiha:2006ke}-\cite{Morales:2006gm}.
 It is also suggested in  \cite{Kallosh:2006bt} that
physical reason for the attractor mechanism is in existence of a
\emph{throat geometry} near the horizon of an extremal black hole.
This throat causes that spatial distance between an arbitrary point
near the horizon to the horizon of an extremal black hole diverges.
In the non-extremal case there is no such throat and this distance
is finite. In \cite{Kallosh:2006bt} it is argued  that this infinite
distance makes the information of the spatial   infinity be lost
when one  approaches  the horizon. Hence due to this infinite
separation the parameters of the horizon become independent of the
asymptotic values. The power of the entropy function method is in
using this throat geometry in the definition of the  extremal black
hole. In \cite{Bardeen:1999px}, existence of the  throat for the
extremal rotating black holes is shown, therefor it seems that the
throat geometry is   a property of all known extremal black holes
 leading to the attractor mechanism.

In the proof of the attractor    mechanism for non-supersymmetric
spherically symmetric charged black holes in  \cite{Kallosh:2006bt},
the double-horizon property of the extremal black holes is used to
reduce the equations of motion for scalar fields to equations which
gives the value of scalars at the horizon without any need to
asymptotic information, more than this, they have shown that the
infinite throat of an extremal black hole is a consequence of the
double-horizon condition. Therefore it seems logical that imposing
the double-horizon conditions on the equations of motion  leads  to
the same results of the  entropy function method does. In this work
we investigate this point.

It is shown that imposing the double-horizon condition plus certain
regularity of the physical properties at the horizon  converts
equations of motion at the horizon to a system of equations which
decouple from the bulk. This decoupling occurs as all r-derivative
terms in the equations of motions disappear at the horizon . Role of
double-horizon becomes completely obvious. For distinct horizons
this decoupling is not achieved and dynamics of the fields at the
horizon  , including the scalar moduli do not decouple from the
bulk.

We show  our statement  for both rotating and non-rotating
electrically charged black holes for  general two-derivative
theories  of gravity which couple to a number of    scalars and
$U(1)$ gauge fields.

For non-rotating case e.o.m at the horizon are converted to a system
of algebraic equations which  can be solved  without using any other
information of the bulk.  The final equations which determine field
configuration of the horizon are very similar to equation which come
from entropy function method.

For rotating black holes equation at the horizon are not algebraic,
they are differential equations. Like non-rotating case the
equations are decoupled from the bulk.   Solutions to these
equations need
 boundary conditions. These boundary conditions are values of
the fields at the poles of the horizon. Necessity of this boundary
values makes it difficult to immediately deduce the attractor
mechanism for rotating black holes.

Adding higher order $f(R)$ corrections   does not change the
decoupling of the horizon from the bulk for double-horizon black
holes.This point extends our result to this class of theories.

\section{ Decoupling of Dynamics at the Horizon and the Double-Horizon Condition }
In this section we  explore the role of the  double-horizon
condition  in the decoupling  of the dynamics of the fields at the
horizon from the rest of the  space.

We consider a theory of gravity with scalars and $U(1)$ gauge
fields. For simplicity  the  theory with only one scalar and one
gauge field is studied. Generalizations to theories with more scalar
and gauge fields are straightforward.  The action which describes
this theory is given by
\begin{equation}\label{3.1}
   S=\frac{1}{\kappa^2}\int d^4 x
    \sqrt{-G}\big(R-2\partial_\mu\Phi\partial^\mu\Phi-f(\Phi)F_{\mu\nu}F^{\mu\nu}\big)
\end{equation}

After studying this model we show that our analysis will work for
more complicated actions which involves different functions of $R$
 in particular $R^n$ gravity and theories with cosmological constant.

Equations of motion are
\begin{equation}\label{3.2}
    R_{\mu\nu}-2\partial_\mu\Phi\partial_\nu\Phi=f(\Phi)(2F_{\mu\lambda}F_\nu\,^\lambda-\frac{1}{2}G_{\mu\nu}F_{\kappa\lambda}F^{\kappa\lambda})
\end{equation}
\begin{equation}\label{3.3}
    \frac{1}{\sqrt{-G}}\partial_\mu(\sqrt{-G}\partial^\mu\Phi)=\frac{1}{4}\frac{\delta
f(\Phi)}{\delta\Phi}F_{\mu\nu}F^{\mu\nu}
\end{equation}
\begin{equation}\label{3.4}
    \partial_\mu(\sqrt{-G}f(\Phi)F^{\mu\nu})=0
\end{equation}
 The ansatz  we consider in the following subsection describes  different black holes of this theory,
  including the
 electrically  charged , both rotating
and non rotating .

\subsection{Non-rotating electrically charged black holes}
The  ansatz for the fields of a non-rotating black hole with only
electrical charge is
\begin{equation}\label{3.5}
    ds^2=-\frac{(r-r_+)(r-r_-)}{A(r)}dt^2+\frac{A(r)}{(r-r_+)(r-r_-)}dr^2+B(r)\Big(d\theta
^2+sin^2(\theta)d\phi^2\Big)
\end{equation}
\begin{equation}\label{3.6}
    F_{rt}=e(r)\qquad\Phi=\Phi(r)
\end{equation}
where $r_+$ and $r_-$ are the radii of the outer and the inner
horizons. $A(r)$, $B(r)$ and $e(r)$ the radial component of the
electric field, are non-zero regular functions. $\Phi(r)$  the
scalar field is also finite at the horizon. For brevity  we show the
value of these functions at the horizon by $A$, $B$, $e$ and $\Phi$.
This form of the metric, scalar and gauge field are consequences of
the spherical symmetry.

On  the horizon we  have
\begin{equation}\label{3.7}
    F\equiv F_{\mu\nu}F^{\mu\nu}\Bigg\vert_{r=r_+}\!\!\!\!\!\!\!\!\!\!=-2e^2
\end{equation}

Allowing the two horizons merge in the equations of motion and using
(\ref{3.5})-(\ref{3.7}), and (\ref{3.3}) on  the outer horizon we
obtain,
\begin{equation}\label{3.8}
    \frac{1}{A}(r_+-r_-)(\frac{\partial}{\partial
    r}\Phi(r))\Bigg\vert_{r=r_+}\!\!\!\!\!\!\!\!\!\!=-\frac{1}{2}e^2\frac{\delta f}{\delta\Phi}\Bigg\vert_{r=r_+}
\end{equation}
The left hand side of this equation  vanishes when we apply the
double-horizon assumption  i.e. $r_+=r_-$ and finiteness of $\Phi$ ,
\begin{equation}\label{3.9}
    \frac{\delta f}{\delta\Phi}\Bigg\vert_{r=r_+}\!\!\!\!\!\!\!\!\!\!=0
\end{equation}
Whatever  the behavior of the fields  at the infinity   , the  value
of the scalar field must satisfy this equation which can be solved
without any other information related to the bulk. It is even
independent of the charge of the black hole. This is an attractor
equation for a non rotating charged black hole in which the
 role of double horizon in removing the r- derivative of $\Phi(r)$
 and hence
independence from the $\Phi $ in the bulk of the space in particular
far distance is obvious.

This equation is in agreement with \cite{Goldstein:2005hq}. It is
the  simplified form of the equation
\begin{equation}\label{effective}
    \frac{\delta V_{eff}}{\delta \Phi}=0
\end{equation}
that  has been used in \cite{Goldstein:2005hq} to obtain the value
of $\Phi$ on the horizon. However, there is a difference, in
\cite{Goldstein:2005hq}, (\ref{effective}) has been  imposed as a
condition for attractor mechanism but we have derived it by imposing
double-horizon condition on the e.o.m. Therefore double-horizon
condition forces  this equation and is a sufficient condition for
the attractor mechanism. Our further analysis shows its necessity.

 Next we consider the effect of the double-horizon on the  other
equations. If we write equation (\ref{3.2}) at the horizon, the
double-horizon condition  converts  it to
\begin{equation}\label{e1}
    f(\Phi)Ae^2=1
\end{equation}
\begin{equation}\label{e2}
    f(\Phi)Be^2=1
\end{equation}

 Using (\ref{3.5}) and (\ref{3.6}) to simplify the equation (\ref{3.4}) ,we
obtain ;
\begin{equation}\label{e3}
    \frac{d}{dr}\Big(sin\theta B(r)f(\Phi)e(r)\Big)=0
\end{equation}
 The double-horizon condition does not have any effect on this
 equation and thus there is no mechanism to remove the
 derivative with respect to   radial directions .
 However, this equation introduces a conserved quantity which for our
case is the charge of the black hole. By integrating both sides of
this equation we get
\begin{equation}
    \frac{d}{dr}\Bigg(\int_0^{2\pi}d\phi\int_0^\pi d\theta
    sin\theta B(r)f(\Phi)e(r)\Bigg)=0
\end{equation}
Hence
\begin{equation}\label{e4}
    \int_0^{2\pi}d\phi\int_0^\pi d\theta\,
    sin\theta B(r)f(\Phi)e(r)=\pi Q
\end{equation}
 where $Q$ is the charge of the black hole. This equation does not
 have any dependence on radial coordinate , thus at the horizon we
 have
\begin{equation}\label{e5}
       Bf(\Phi)\,e=\frac{1}{4} Q
\end{equation}

Equations (\ref{3.9}), (\ref{e1}), (\ref{e2}) and (\ref{e5}) are
attractor equations. They completely determine all the  information
at the  horizon without any reference  to the information in the
bulk.

Entropy of this black hole  is equal to $\frac{A_r}{4}$ where $A_r$
is the area of the horizon. Using (\ref{e2}) and (\ref{e5}) ,
entropy is given by
\begin{equation}\label{entropy of non-rotating}
    \varepsilon=\frac{\pi}{16}\frac{Q^2}{f(\Phi)}
\end{equation}

As we see, the double-horizon condition is just sufficient to remove
 the r-derivative terms   in the
equations which originate from the variation of metric and scalar
field. However the equation of the gauge field is not decoupled by
double-horizon condition. But   the charge of the black hole is
sufficient to solve this equation and specify the value of the
electromagnetic fields at  the horizon. Therefore by knowing the
charge of the double-horizon black holes we can extract every
information about the horizon. If these equations have unique
solutions then it results a \emph{generalized attractor mechanism}
which states that the value of the  fields at the horizon of a
double-horizon black hole is independent of the asymptotic values.
For the case where the equations have  more than one solution the
conclusion is not immediate like \cite{Goldstein:2005hq} where we
must consider dynamics of the fields near the horizon  to determine
which of these solutions is reached. In this case the black hole may
assume different phases corresponding to the different values of the
scalar moduli.


\subsection{Rotating Electrically Charged Black Holes}

We consider  the same theory which is described by the action
(\ref{3.1}). Equations of motion are (\ref{3.2})-(\ref{3.4}). We
shall  investigate a rotating black hole in this theory. We take the
angular momentum of this black hole to be along the   $z$ direction.
Our ansatz for the metric of the  rotating black hole is
\begin{equation}\label{3.25}
 \begin{split}
 ds^2=&B(r,\theta)\Bigg(-\frac{S(r)}{A(r)-a^2\, sin^2\theta\, S(r)}dt^2+\frac{1}{S(r)}dr^2+d\theta^2\Bigg)
 \cr
   &+sin^2(\theta)\frac{A(r)-a^2\, sin^2\theta\, S(r)}{B(r,\theta)}\Bigg(d\phi-\frac{C(r)-E(r,\theta)S(r)}{A(r)-a^2\, sin^2\theta\, S(r)}dt\Bigg)^2
  \end{split}
\end{equation}
 where  $a=\frac{J}{M}$, $J$ is the angular momentum , $M$
is the mass of the black hole and we have,
\begin{equation}\label{3.11}
    S(r)\equiv(r-r_+)(r-r_-)
\end{equation}
 where $r_+$ and $r_-$ are respectively radii of the outer and the inner
horizons and functions $A(r)$, $B(r,\theta)$,   $C(r)$ and
$E(r,\theta)$ are non-zero  regular functions at the horizon. We
show their value  at the outer horizon by $A$, $B(\theta)$, $C$ and
$E(\theta)$ which are regular function at the horizon.

Unlike the non-rotating case the scalar field $\Phi$  depends   on
$r$ and $\theta$ coordinates:
\begin{equation}\label{scalar}
    \Phi=\Phi(r,\theta)
\end{equation}
Independence from $\phi$-coordinate  results from azimuthal
symmetry.

It is assumed that our black hole is  only electrically charged.
 Our ansatz for the gauge field is
 \begin{equation}\label{gauge field}
    A_t= b(r,\theta)\qquad A_\phi=-a\,sin^2\theta \,b(r,\theta)
\end{equation}
 Therefore only $F_{rt}$, $F_{r\phi}$,
$F_{\theta t}$ and $F_{\theta\phi}$ of $F_{\mu\nu}$ tensor  are
non-zero:
\begin{equation}\label{filed streng r}
    F_{rt}=\frac{\partial}{\partial r}A_t\equiv e(r,\theta)\qquad
    F_{r\phi}=-asin^2\theta\, e(r,\theta)
\end{equation}
\begin{equation}\label{field streng theta}
    F_{\theta t}=\frac{\partial}{\partial \theta}\,b(r,\theta)\qquad
    F_{\theta\phi }=\frac{\partial}{\partial
    \theta}\big(-asin^2\theta\,
    b(r,\theta)\big)
\end{equation}
We show the value of $b(r,\theta)$ and $e(r,\theta)$ at the horizon
by $b(\theta)$ and $e(\theta)$.

Our ansatz for the field configuration of a rotating black hole is
compatible with the Kerr-Newman solution and  the solutions of
\cite{Sen:1994eb}\footnote{Solutions of \cite{Sen:1994eb} are
completely general and consist of  black holes of more than one
charge but for the case that black hole has only one kind of charge,
our ansatz is compatible with those solutions. }.

By using symmetry consideration we can deduce that at the horizon
\begin{equation}\label{boundary value f B}
    B(\theta=0)=B(\theta=\pi)
\end{equation}
\begin{equation}\label{boundary value f Phi}
    \Phi(\theta=0)=\Phi(\theta=\pi)
\end{equation}
\begin{equation}\label{boundary value f b}
    b(\theta=0)=b(\theta=\pi)
\end{equation}

As a physical condition we demand that all the physical quantities
to be finite on  the horizon. Using (\ref{3.25}), (\ref{filed streng
r}) and (\ref{field streng theta}) we get
\begin{equation}\label{FmunuFmunu}
\begin{split}
    F_{\mu\nu}F^{\mu\nu}=-&\frac{2}{B(r,\theta)}\Bigg(\frac{A(r)-a^2\,sin^2\theta S(r)}{B(r,\theta)}F_{rt}^2+\frac{C(r)^2}{B(r,\theta)\big(A(r)-a^2\,sin^2\theta S(r)\big)}F_{r\phi}^2\cr+&2\frac{C(r)-S(r)E(r,\theta)}{B(r,\theta)}F_{rt}F_{r\phi}\Bigg)
    -\frac{2}{B(r,\theta)S(r)}\Bigg(\frac{A(r)-a^2\,sin^2\theta S(r)}{B(r,\theta)}F_{\theta t}^2\cr+&\frac{C(r)^2}{B(r,\theta)\big(A(r)-a^2\,sin^2\theta S(r)\big)}F_{\theta\phi}^2+2\frac{C(r)-S(r)E(r,\theta)}{B(r,\theta)}F_{\theta t}F_{\theta\phi}\Bigg)
    \cr+&\frac{2}{B(r,\theta)}\Big(S(r)F_{r\phi}^2+F_{\theta\phi}^2\Big)\Bigg(\frac{B(r,\theta)^2-2sin^2\theta\,C(r)E(r,\theta)}{sin^2\theta \,B(r,\theta)\big(A(r)-a^2\,sin^2\theta S(r)\big)}\cr-&\frac{S(r)E(r,\theta)^2}{B(r,\theta)\big(A(r)-a^2\,sin^2\theta
    S(r)\big)}\Bigg)
\end{split}
\end{equation}
 Finiteness  of $F_{\mu\nu}F^{\mu\nu}$
at the horizon requires  that
\begin{equation}\label{finitness of field strenght}
    F_{\theta t}+\frac{C}{A}F_{\theta\phi}=0
\end{equation}
 This is a reduced form for the following
equation:
\begin{equation}\label{genera ,finitness of field strenght}
    F_{\theta t}+\frac{C(r)-E(r,\theta)S(r)}{A(r)-a^2\, sin^2\theta\, S(r)}F_{\theta\phi}=V(r,\theta)S(r)
\end{equation}
where $V(r,\theta)$ is a non-zero finite function at the
horizon.\newline
 An immediate result of this equation is that
\begin{equation}\label{equatin for F}
    F\equiv F_{\mu\nu}F^{\mu\nu}\Bigg\vert_{r=r_+}\!\!\!\!\!\!\!\!\!\!\!\!=\frac{2}{A\,sin^2\theta}\Big(1+\frac{a^2\, sin^4\,\theta C^2}{A\,B(\theta)^2}\Big)F_{\theta\phi}^2-\frac{2A}{B(\theta)^2}\Big(F_{rt}+\frac{C}{A}F_{r\phi}\Big)^2
\end{equation}

 Using (\ref{genera ,finitness of field strenght}) and imposing the
double-horizon condition we have

\begin{equation}\label{3.33}
   \alpha\equiv\frac{\partial}{\partial r}\frac{
    C(r)}{A(r)}\Bigg\vert_{r=r_h}\!\!\!\!\!\!\!\!\!=-\frac{\partial}{\partial r}\Bigg(\frac{F_{\theta t}}{F_{\theta \phi}}\Bigg)\Bigg\vert_{r=r_h}
\end{equation}
where  $\alpha$ is a constant. We can calculate it by using only the
information at the horizon, because

\begin{equation}\label{3.34}
    \frac{\partial}{\partial r}\frac{F_{\theta t}}{F_{\theta \phi}}=\frac{1}{F_{\theta
    \phi}^2}\bigg(F_{\theta \phi}\frac{\partial}{\partial r}\frac{\partial}{\partial \theta}\big(A_t\big)-F_{\theta t}\frac{\partial}{\partial r}\frac{\partial}{\partial \theta}\big(A_\phi\big)\bigg)
\end{equation}
by exchanging order of $\frac{\partial}{\partial r}$ and $
\frac{\partial}{\partial \theta}$ and using (\ref{finitness of
field strenght})  we get
\begin{equation}\label{3.35}
\begin{split}
    \alpha=&-\frac{1}{F_{\theta
    \phi}^2}\bigg(F_{\theta \phi}\frac{d}{d
    \theta}F_{rt}-F_{\theta t}\frac{d}{d
    \theta}F_{r\phi}\bigg)\cr
    =&-\frac{1}{F_{\theta \phi}}\frac{d}{d
    \theta}\Big(F_{rt}+\frac{C}{A} F_{r\phi}\Big)
 \end{split}
\end{equation}
For Kerr-Newman solution $\alpha$ is given by
\begin{equation}
   \alpha=\frac{-2J}{(M^2+a^2)^2}
\end{equation}

After deriving the consequences  of the finiteness of the physical
quantities at the horizon , we consider the  role of the
double-horizon condition on the form of the e.o.m at the horizon.
Our claim is that , this condition converts the equations at the
horizon to a system of equations which are completely independent of
 the radial coordinate.

We start from the equation (\ref{3.3}). At the outer horizon it
reads;
\begin{equation}\label{3.26}
   \frac{1}{\sqrt{-G}}\frac{\partial}{\partial r}\left(\sqrt{-G}\frac{S(r)}{B(r,\theta)}\frac{\partial \Phi}{\partial
   r}\right)\Bigg\vert_{r=r_+}\!\!\!\!\!\!\!+\frac{1}{\sqrt{-G}}\frac{\partial}{\partial \theta}\left(\sqrt{-G}\frac{1}{B(r,\theta)}\frac{\partial \Phi}{\partial
   \theta}\right)\Bigg\vert_{r=r_+}\!\!\!\!\!\!\!=\frac{1}{4}\frac{\delta f}{\delta \Phi}F
\end{equation}

Since $S(r)$ and $\frac{d }{dr}S(r)$ are zero at the horizon  the
double-horizon condition forces  the first term of the L.H.S of this
equation vanish. Thus we obtain
\begin{equation}\label{3.27}
    \frac{1}{\sqrt{-G}}\frac{d}{d  \theta}\left(\sqrt{-G}\frac{1}{B(\theta)}\frac{d
    \Phi}{d
   \theta}\right)=\frac{1}{4}\frac{\delta f}{\delta \Phi}F
\end{equation}
whereupon  using (\ref{3.25}),gives
\begin{equation}\label{3.28}
    G=-B(\theta)^2sin^2\theta
\end{equation}
Substituting (\ref{3.28}) and (\ref{equatin for F}) in (\ref{3.27})
gives the following equation:
\begin{equation}\label{scalar final simplifyed eq}
    \frac{1}{sin\theta}\frac{d}{d  \theta}\left(sin\theta\frac{d
    \Phi}{d
   \theta}\right)=\frac{1}{2}\frac{\delta f}{\delta
   \Phi}\Bigg(\frac{B(\theta)}{A\,sin^2\theta}\Big(1+\frac{a^2\,sin^4\,\theta C^2}{A\,B(\theta)^2}\Big)F_{\theta\phi}^2-\frac{A}{B(\theta)}\Big(F_{rt}+\frac{C}{A}F_{r\phi}\Big)^2\Bigg)
\end{equation}
which  can be written as
\begin{equation}\label{scalar final simplifyed eq1}
    \frac{1}{sin\theta}\frac{d}{d  \theta}\left(sin\theta\frac{d
    \Phi}{d
   \theta}\right)=\frac{\delta V_{eff}}{\delta\Phi}
\end{equation}
with $V_{eff}$  defined to be
\begin{equation}\label{Veff}
    V_{eff}=\frac{1}{2}f(\Phi)\frac{A}{B(\theta)}\Bigg(\Big(\frac{B(\theta)}{A\,sin\theta}F_{\theta\phi}\Big)^2\Big(1+\frac{a^2\,sin^4\,\theta
    C^2}{A\,B(\theta)^2}\Big)-\Big(F_{rt}+\frac{C}{A}F_{r\phi}\Big)^2\Bigg)
\end{equation} ,

 We see again that the double-horizon condition is responsible for  the
 removal of the
r-derivative. This is one of the attractor equations but unlike the
non-rotating case where the  scalar decouples from the  other
fields, determination of scalar field at the horizon relies on the
knowledge of the  metric and gauge field strength albeit restricted
to the horizon. Concluding the  attractor mechanism for the scalar
field needs the demonstration of  the independence of the geometry
and the gauge field strength at the horizon from the asymptotic
value of the scalar field. We show this point by investigating the
other equations of motion.

 Another important difference with the non-rotating case is that
(\ref{scalar final simplifyed eq1}) is a differential equation, thus
complete determination of scalar field at the horizon needs imposing
certain  boundary conditions.  These boundary conditions are only at
the horizon but if some  attractor mechanism is at work it is
required
 that the boundary conditions be independent of the asymptotic
values. Proof of this point needs more considerations.

 Now we consider imposing the double-horizon condition on the
 gravity part of the
equations at the horizon . A straightforward calculation using  the
metric (\ref{3.25}) shows that the non-zero components of the Ricci
tensor are $R_{rr}$, $R_{r\theta}$, $R_{\theta\theta}$, $R_{tt}$,
$R_{\phi\phi}$ and $R_{t\phi}$.

At the horizon, the equation (\ref{3.2}) for the $R_{rr}$ $R_{tt}$,
$R_{\phi\phi}$ and
 $R_{t\phi}$ components reduces to
\begin{equation}\label{eom for Rrr}
   1+\frac{1}{2}\frac{d}{d
   \theta}\Big(\frac{1}{B(\theta)}\frac{d}{d\theta}B(\theta)\Big)+\frac{cot\,\theta}{2B(\theta)}\frac{d}{d\theta}B(\theta)-\frac{1}{2}\alpha^2\frac{A^2}{B(\theta)^2}sin^2\theta=-\frac{A}{B(\theta)}f(\Phi)H(\theta)
\end{equation}
where $\alpha$ is given by (\ref{3.35}) and
\begin{equation}\label{def H}
    H(\theta)=\Bigg(\big(F_{r t}+\frac{C}{A}F_{r\phi}\big)^2+\Big(1-\frac{a^2\,sin^4\,\theta C^2}{A\,B(\theta)^2}\Big)\bigg(\frac{B(\theta)F_{\theta\phi}}{A sin\theta}\bigg)^2\Bigg)
\end{equation}

For the $R_{\theta\theta}$, the equation (\ref{3.2}) at the horizon
simplifies and takes the form,
\begin{equation}\label{eq thetatheta}
    -1+\frac{1}{2B(\theta)}\frac{d^2}{d\,\theta^2}B(\theta)-\frac{3cot\theta}{2B(\theta)}\frac{d}{d\theta}B(\theta)=2\Big(\frac{d}{d\theta}\Phi(\theta)\Big)^2+f(\Phi)\frac{A}{B(\theta)}H(\theta)
\end{equation}

For the $R_{r\theta}$ component, equation (\ref{3.2}) at the horizon
reads as
\begin{equation}\label{Rr theta}
    F_{rt}F_{\theta t}+\frac{C^2}{A^2}F_{r\phi}F_{\theta\phi}+\frac{C}{A}\big(F_{rt}F_{\theta \phi}+F_{r\phi}F_{\theta
    t}\big)=0
\end{equation}
But if we use (\ref{finitness of field strenght}), this equation is
clearly  satisfied. It is obvious that again the double-horizon
condition has removed  all the r-derivative terms in the above
equations.

Now we consider  the equation (\ref{3.4}) from which we obtain ;
\begin{equation}\label{3.46}
    \frac{\partial}{\partial r}\Big(\sqrt{-G}f(\Phi)F^{rt}\Big)+\frac{\partial}{\partial \theta}\Big(\sqrt{-G}f(\Phi)F^{\theta
    t}\Big)=0
\end{equation}
\begin{equation}\label{3.47}
   \frac{\partial}{\partial r}\Big(\sqrt{-G}f(\Phi)F^{r\phi}\Big)+\frac{\partial}{\partial \theta}\Big(\sqrt{-G}f(\Phi)F^{\theta
   \phi}\Big)=0
\end{equation}
  Using (\ref{3.25}), we can write
\begin{equation}\label{3.48}
    F^{r\phi}=\frac{C(r)-E(r,\theta)S(r)}{A(r)-a^2\, sin^2\theta\, S(r)}F^{rt}+\frac{S(r)}{sin^2\theta\,A(r)}F_{r\phi}
\end{equation}
\begin{equation}\label{3.49}
    F^{\theta\phi}=\frac{C(r)-E(r,\theta)S(r)}{A(r)-a^2\, sin^2\theta\, S(r)}F^{\theta t}+\frac{1}{sin^2\theta\,A(r)}F_{\theta\phi}
\end{equation}
 Substituting (\ref{3.48}) and (\ref{3.49}) in (\ref{3.47})   and imposing the double-horizon assumption, at
the horizon we find ;

 \begin{equation} \label{3.50}
  \begin{split}
  &\frac{C}{A} \Bigg(\frac{\partial}{\partial
r}\Big(\sqrt{-G}f(\Phi)F^{rt}\Big)\Bigg\vert_{r=r_+}\!\!\!\!\!\!\!\!\!+\frac{\partial}{\partial
\theta}\Big(\sqrt{-G}f(\Phi)F^{\theta
    t}\Big)\Bigg)+\cr&+\Big(\sqrt{-G}f(\Phi)F^{rt}\Big)\frac{\partial}{\partial
    r}\Big(\frac{C(r)}{A(r)}\Big)\Bigg\vert_{r=r_+}\!\!\!\!\!\!\!\!\!
    +\frac{1}{A}\frac{d}{d
   \theta}\Big(\frac{1}{sin^2(\theta)}\sqrt{-G}f(\Phi)F_{\theta\phi}\Big)=0
\end{split}
\end{equation}
Using (\ref{3.46}) and (\ref{3.33}) in (\ref{3.50}) ,
 we can remove the r-derivatives to obtain,
\begin{equation}\label{3.51}
   \alpha \Big(\sqrt{-G}f(\Phi)F^{rt}\Big)+\frac{1}{A}\frac{d}{d
   \theta}\Big(\frac{1}{sin^2(\theta)}\sqrt{-G}f(\Phi)F_{\theta\phi}\Big)=0
\end{equation}
  Finally substitution of  (\ref{3.28}) and (\ref{3.35}) in this equation
gives;
\begin{equation}\label{final equation from gauge field }
    f(\Phi)^2\frac{d}{d\theta}\Big(F_{rt}+\frac{C}{A}F_{r\phi}\Big)^2+\frac{d}{d\theta}\Big(\frac{f(\Phi)B(\theta)}{A\,sin\theta}F_{\theta\phi}\Big)^2=0
\end{equation}

Integrating the  equation (\ref{3.46}) results ;
\begin{equation}
    \frac{\partial}{\partial r}\Bigg(\int_0^\pi d\theta
    \sqrt{-G}f(\Phi)F^{rt}\Bigg)+\Bigg[\sqrt{-G}f(\Phi)F^{\theta
    t}\Bigg]_{\theta=0}^{\theta=\pi}=0
\end{equation}.
The second term
 vanishes
due to  $sin\theta$ coefficient in $\sqrt{-G}$ and the regularity of
the components of $F_{\mu\nu}$  at the horizon. Hence the first term
introduces a constant quantity $Q$, the charge of the black hole:
\begin{equation}
    \int_0^\pi d\theta
    \sqrt{-G}f(\Phi)F^{rt}=\frac{1}{2}Q
\end{equation}

Integrating (\ref{3.51}) and substituting in this equation  implies
that:
\begin{equation}
    \frac{1}{2}\alpha
    Q=-\frac{1}{A}\Big[\frac{\sqrt{-G}f(\Phi)F_{\theta\phi}}{sin^2\theta }\Big]_{\theta=0}^{\theta=\pi}
\end{equation}
Use of  (\ref{3.35}) and (\ref{3.28}) in this equation gives;
\begin{equation}\label{simplifyed equation of gauge filed e1}
    \frac{d}{d\theta}\Big(F_{rt}+\frac{C}{A}F_{r\phi}\Big)-\frac{\beta}{A}
    F_{\theta\phi}=0
\end{equation}
where
\begin{equation}\label{definition of beta}
    \beta=\Big[\frac{2f(\Phi)B(\theta)F_{\theta\phi
    }}{Q\,sin\theta}\Big]_{\theta=0}^{\theta=\pi}
\end{equation}
which simplifies with the aid of
 (\ref{field streng theta})-(\ref{boundary value f b}) give
\begin{equation}\label{final form of beta}
    \beta=\frac{8a}{Q}f(\Phi)B(\theta)b(\theta)\Big\vert_{\theta=0}
\end{equation}
$\beta$ is a constant and  depends on  the values of the fields at
the poles of the horizon. Values of the  fields at the poles are the
necessary  boundary conditions in our discussion, thus knowing this
value will determine $\beta$. It is notable  that $\beta$ depends on
the parameters of the black hole, for example for Kerr-Newman
solution we have
\begin{equation}
    \beta=2J
\end{equation}
Using (\ref{3.35}) and (\ref{simplifyed equation of gauge filed e1})
we deduce,
\begin{equation}\label{relation alpha and beta}
    \alpha=-\frac{\beta}{A}
\end{equation}

Equations (\ref{finitness of field strenght}), (\ref{scalar final
simplifyed eq1}), (\ref{eom for Rrr}), (\ref{eq thetatheta}),
(\ref{final equation from gauge field }) and (\ref{simplifyed
equation of gauge filed e1}) are a complete set of independent
differential equations which determine $A$, $B(\theta)$, $C$,
$\Phi(\theta)$, $b(\theta)$ and $e(\theta)$ ( we summarize them at
the Appendix  ). These equations are independent of  the radial
direction and the complete solution of them only needs values of the
fields at the poles of the horizon as the boundary condition. These
equations include  the constant $\alpha$, $\beta$, $a$ and $Q$ which
depend on the physical observable of the solution. Only two
quantities are independent.

Independence of the dynamics of the fields at the horizon from the
bulk is a consequence of the double horizon condition but the final
conclusion on the  independence of the values of the fields at the
horizon from asymptotic values (attractor mechanism) needs more
considerations, first  we must investigate the independence of the
boundary conditions ( values of the fields at the poles of the
horizon) from the asymptotic values and second, the solution must be
unique. When there are , more than one solution like non-rotating
case we must consider dynamics of the fields near the  horizon. It
is enough to perform the analysis  near  one of the poles .

\section{Generalization to f(R) Gravities}
The essential role of the double-horizon condition in isolating
equations of motion at the horizon from the bulk  also occurs
 in the theories of gravity with higher order corrections.

 A simple generalization of the  theory  considered so far, is  addition of cosmological constant. The
action has the form
\begin{equation}\label{4.1}
    S=\frac{1}{\kappa^2}\int d^4 x
    \sqrt{-G}\big(R-2\Lambda-2\partial_\mu\Phi\partial^\mu\Phi-f(\Phi)F_{\mu\nu}F^{\mu\nu}\big)
\end{equation}
where $\Lambda$ is the cosmological constant.

The equation of motion from variation of scalar and gauge field are
unchanged and they are the same as (\ref{3.3}) and (\ref{3.4}). The
equation of motion from variation of metric is now given by,
\begin{equation}\label{4.2}
    R_{\mu\nu}-\Lambda G{\mu\nu}-2\partial_\mu\Phi\partial_\nu\Phi=f(\Phi)(2F_{\mu\lambda}F_\nu\,^\lambda-\frac{1}{2}G_{\mu\nu}F_{\kappa\lambda}F^{\kappa\lambda})
\end{equation}

It is not difficult to see that by taking the ansatzs (\ref{3.25}),
(\ref{scalar}) and (\ref{gauge field})  for the fields of the black
hole solutions of this theory,  we only have to slightly modify the
equation (\ref{eom for Rrr}) and (\ref{eq thetatheta}) by solely
adding the term  $\Lambda B(\theta)$. Independence of the dynamics
of the fields at the horizon from the bulk in this case is then
clear.

Another example is the $R^n$ gravities. The action is given by
\begin{equation}\label{action of R^n }
    S=\frac{1}{\kappa^2}\int d^4 x
    \sqrt{-G}\big(R+\gamma R^n-2\partial_\mu\Phi\partial^\mu\Phi-f(\Phi)F_{\mu\nu}F^{\mu\nu}\big)
\end{equation}

The equations of motion from the variation of metric are,
\begin{equation}\label{R^n metric equation}
    R_{\mu\nu}+\gamma R^{n-1}(2R_{\mu\nu}-\frac{n-1}{2}RG_{\mu\nu})-2\partial_\mu\Phi\partial_\nu\Phi=f(\Phi)(2F_{\mu\lambda}F_\nu\,^\lambda-\frac{1}{2}G_{\mu\nu}F_{\kappa\lambda}F^{\kappa\lambda})
\end{equation}
The equations of motion from the variation of the scalar and gauge
field are unchanged and the same as (\ref{3.3}) and (\ref{3.4}).

The extra term in the equation (\ref{R^n metric equation}) adds the
following term to the equations (\ref{eom for Rrr}):
\begin{equation}\label{additinal term to Rrr}
   \gamma I_{1}(\theta)^{n-1}\Big(I_{2}(\theta)-\frac{A^2sin^2\theta}{B(\theta)^2}\,\alpha^2+\frac{n-1}{2}I_{1}(\theta)B(\theta)\Big)
\end{equation}
where
\begin{equation}\label{I1}
    I_1(\theta)=-\frac{A^2sin^2\theta}{2B(\theta)^3}\,\alpha^2+\frac{1}{2B(\theta)}\frac{d}{d\theta}\Big(\frac{1}{B(\theta)}\frac{d}{d\theta}B(\theta)\Big)+\frac{1}{B(\theta)^2}\frac{d^2}{d\theta^2}B(\theta)-\frac{cot\theta}{B(\theta)^2}\frac{d}{d\theta}B(\theta)
\end{equation}
\begin{equation}\label{I2}
    I_2(\theta)=2+\frac{d}{d
   \theta}\Big(\frac{1}{B(\theta)}\frac{d}{d\theta}B(\theta)\Big)+\frac{cot\,\theta}{B(\theta)}\frac{d}{d\theta}B(\theta)
\end{equation}

For the equation (\ref{eq thetatheta}) the additional term is
\begin{equation}\label{additinal term for Rthetetheta}
    \gamma I_{1}(\theta)^{n-1}\Big(I_{3}(\theta)-\frac{n-1}{2}I_{1}(\theta)B(\theta)\Big)
\end{equation}
where
\begin{equation}\label{I3}
    I_3(\theta)=-2+\frac{1}{B(\theta)}\frac{d^2}{d\,\theta^2}B(\theta)-\frac{3cot\theta}{B(\theta)}\frac{d}{d\theta}B(\theta)
\end{equation}

Other equations are unchanged. It is clear that our claim remains
valid and there is no  r-derivative terms in the  equations of
motion. Hence the  dynamics of the fields at the horizon is
independent of the bulk. This shows that for every well-defined
function of $R$ which is expandable   in terms of the powers of $R$,
this decoupling
 occurs. Therefore the double-horizon condition  is a necessary condition for
attractor mechanism in $f(R)$ gravities and is responsible for the
decoupling of the dynamics at the horizon from the rest of the
space.


%

\section{Conclusion}
Our analysis  shows that the double-horizon condition in the
theories we have considered, namely $f(R)$ theories, has capability
of decoupling the dynamics at the horizon of the black hole
solutions . This isolation is done by removal of the  radial
coordinate derivatives in the equations of motion. For non-rotating
black holes, e.o.m at the horizon are completely algebraic and it is
possible to solve them without any other information.  If the
solution is unique , it leads to the attractor mechanism , if not ,
the resolution needs detail considerations. One must analyze the
stability of the solution with respect to changes in the bulk.

 For the rotating black hole, e.o.m are differential equations
 of functions of  $\theta$ coordinate. Solving them needs boundary
 conditions.
 These boundary conditions are values of the fields at the poles of
 the black hole.  Necessity of these boundary
condition makes it difficult to immediately arrive at the conclusion
about the attractor mechanism. The attractor mechanism needs the
investigation of the  independence of the value of the fields at the
poles from asymptotic values. Since the equations we are considering
are on a compact manifold ( sphere ) and are non-linear, single
valuedness may act as boundary condition. Therefore it is plausible
that consistency condition will dictate a certain class of boundary
condition(s) on the solutions. Moreover if the solution to equations
of motion at the horizon is not unique stability analysis is
required to differentiate between different configurations  . Again
detailed analysis is needed like what has been done for non-rotating
case in \cite{Goldstein:2005hq}. In any case solutions to the set of
the equations obtained, specify the only possible field
configurations and will limit the moduli at the horizon. Any field
configuration must assume one of these forms as its limit at the
horizon. Such questions are under investigation by the authors.

 We have considered  black holes with only electric charge but by
 changing our ansatz for the gauge fields, it is possible to investigate
 black holes with magnetic charge   too.
 It is possible to generalize this method for
 black holes with more than on kind of charge and the theories of
 gravity with more complicated higher order corrections.

Our method clearly shows why the attractor mechanism does not exist
for black holes with distinct horizons. In these cases the  dynamics
of the fields at the outer horizon are not independent of the bulk
and so it is not possible to isolate the horizon from the rest of
the space. All known extremal black holes have double-horizon, thus
our analysis includes this class of  black holes . However, finding
clear relation between extremity and double-horizon limit  needs
further investigations.

If we accept the double-horizon condition as the necessary condition
for the attractor mechanism, role of supersymmetry will be
negligible and one can conclude that it is not the supersymmetry
which fixes the moduli at the horizon. Fixing moduli is a
consequence of infinite throat which makes the time to reach the
horizon from any point in the bulk to diverge and become  infinite.
Our analysis supports this point for generic black holes by direct
investigation of the dynamics of the fields at the horizon. \\ \\
\\\textbf {Note Added :}

  During the time that we were preparing this paper, the attractor mechanism
   has been investigated for rotating black holes by entropy function method
 \cite{Astefanesei:2006dd}. Our results is in completely agreement
 with their results for the case of two-derivative black holes which
 they have studied in that paper.

\section*{Acknowledgments}
We would like to thank F. Ardalan, M. M. Sheikh Jabbari and M.
Alishahiha for their useful comments and discussions. We would also
like to thank the Iranian chapter of TWAS for partial support.
\section*{Appendix}
\appendix
\section{Attractor equations for charged rotating black holes}
We collect the equations of the section (2.2) in this section:

- Metric of the rotating black hole:
\begin{equation}
 \begin{split}
 ds^2=&B(r,\theta)\Bigg(-\frac{S(r)}{A(r)-a^2\, sin^2\theta\, S(r)}dt^2+\frac{1}{S(r)}dr^2+d\theta^2\Bigg)
 \cr
   &+sin^2(\theta)\frac{A(r)-a^2\, sin^2\theta\, S(r)}{B(r,\theta)}\Bigg(d\phi-\frac{C(r)-E(r,\theta)S(r)}{A(r)-a^2\, sin^2\theta\, S(r)}dt\Bigg)^2
  \end{split}
\end{equation}

- Condition for the finiteness of the $F_{\mu\nu}F^{\mu\nu}$ at the
horizon:
\begin{equation}
    F_{\theta t}+\frac{C}{A}F_{\theta\phi}=0
\end{equation}

-Equations of motion at the horizon after imposing double horizon
condition:
\begin{equation}
    \frac{1}{sin\theta}\frac{d}{d  \theta}\left(sin\theta\frac{d
    \Phi}{d
   \theta}\right)=\frac{\delta V_{eff}}{\delta\Phi}
\end{equation}
\begin{equation}
   1+\frac{1}{2}\frac{d}{d
   \theta}\Big(\frac{1}{B(\theta)}\frac{d}{d\theta}B(\theta)\Big)+\frac{cot\,\theta}{2B(\theta)}\frac{d}{d\theta}B(\theta)-\frac{1}{2}\alpha^2\frac{A^2}{B(\theta)^2}sin^2\theta=-\frac{A}{B(\theta)}f(\Phi)H(\theta)
\end{equation}
\begin{equation}
    -1+\frac{1}{2B(\theta)}\frac{d^2}{d\,\theta^2}B(\theta)-\frac{3cot\theta}{2B(\theta)}\frac{d}{d\theta}B(\theta)=2\Big(\frac{d}{d\theta}\Phi(\theta)\Big)^2+f(\Phi)\frac{A}{B(\theta)}H(\theta)
\end{equation}
\begin{equation}
    F_{rt}F_{\theta t}+\frac{C^2}{A^2}F_{r\phi}F_{\theta\phi}+\frac{C}{A}\big(F_{rt}F_{\theta \phi}+F_{r\phi}F_{\theta
    t}\big)=0
\end{equation}
\begin{equation}
    f(\Phi)^2\frac{d}{d\theta}\Big(F_{rt}+\frac{C}{A}F_{r\phi}\Big)^2+\frac{d}{d\theta}\Big(\frac{f(\Phi)B(\theta)}{A\,sin\theta}F_{\theta\phi}\Big)^2=0
\end{equation}
\begin{equation}
    \frac{d}{d\theta}\Big(F_{rt}+\frac{C}{A}F_{r\phi}\Big)-\frac{\beta}{A}
    F_{\theta\phi}=0
\end{equation}
where
\begin{equation}
    V_{eff}=\frac{1}{2}f(\Phi)\frac{A}{B(\theta)}\Bigg(\Big(\frac{B(\theta)}{A\,sin\theta}F_{\theta\phi}\Big)^2\Big(1+\frac{a^2\,sin^4\,\theta C^2}{A\,B(\theta)^2}\Big)-\Big(F_{rt}+\frac{C}{A}F_{r\phi}\Big)^2\Bigg)
\end{equation}
\begin{equation}
    H(\theta)=\Bigg(\big(F_{r t}+\frac{C}{A}F_{r\phi}\big)^2+\Big(1-\frac{a^2\,sin^4\,\theta C^2}{A\,B(\theta)^2}\Big)\bigg(\frac{B(\theta)F_{\theta\phi}}{A sin\theta}\bigg)^2\Bigg)
\end{equation}
\begin{equation}
    \beta=\frac{8a}{Q}f(\Phi)B(\theta)b(\theta)\Big\vert_{\theta=0}
\end{equation}
\begin{equation}
    \alpha=-\frac{\beta}{A}
 \end{equation}
Note that these equations are independent of our ansatz for the
gauge field.
%
\end{document}